\documentclass[useAMS,usenatbib]{mn2e}
\usepackage{epsfig,lscape} 
\usepackage[dvipdfm]{hyperref} 
\usepackage{natbib}
\newcommand{\dfn}{\hbox{$D_{\rm n}$(4000)}}
\newcommand{\ha}{\hbox{H$\alpha$}}
\newcommand{\hb}{\hbox{H$\beta$}}
\newcommand{\oii}{\hbox{[O\,{\sc ii}]}}
\newcommand{\nii}{\hbox{[N\,{\sc ii}]}}
\newcommand{\oiii}{\hbox{[O\,{\sc iii}]}}
\newcommand{\hi}{\hbox{H\,{\sc i}}}
\newcommand{\hii}{\hbox{H\,{\sc ii}}}

\title[Mass metallicity relation of LBAs]
{The mass--metallicity relation of Lyman-break analogues and its 
dependence on galaxy properties }
\author[J. H. Lian et al.]
{J. H. Lian,\thanks{E-mail:ljhhw@mail.ustc.edu.cn (JHL);
xkong@ustc.edu.cn(XK)} 
J. R. Li , \thanks{J. H. Lian and J. R. Li contributed equally to this work}
W. Yan and
X. Kong \\
Department of Astronomy, University of Science and Technology of China, Hefei 230026, China\\
Key Laboratory for Research in Galaxies and Cosmology, Chinese Academy of Sciences, Hefei 230026, China}

\begin{document}
\maketitle

\begin{abstract}
We investigate the mass--metallicity relation and its dependence on galaxy 
physical properties with a sample of 703 Lyman-break analogues (LBAs) in 
local Universe, which have similar properties to high redshift star-forming
 galaxies. The sample is selected according to $\ha$ luminosity, 
$L(\ha)>10^{41.8}\,{\rm erg\,s^{-1}}$, and surface brightness, 
$I(\ha)>10^{40.5}\,{\rm erg\,s^{-1}\,kpc^{-2}}$, criteria. 
The mass--metallicity relation of LBAs harmoniously agrees with that of 
star-forming galaxies at $z \sim$ 1.4--1.7 in stellar mass range of 
$10^{8.5}M_{\odot}<M_{*}<10^{11}M_{\odot}$.
The relation between stellar mass, metallicity and star formation rate of 
our sample is roughly consistent with the local fundamental metallicity 
relation. We find that the 
mass--metallicity relation shows a strong correlation with the 4000\AA\, 
break; galaxies with higher 4000\AA\, break typically have higher
metallicity at a fixed mass, by 0.06 dex in average. 
This trend is independent of the methodology of metallicity.  We also use the metallicity 
estimated by $T_{\rm e}$-method to confirm it.  
The scatter in mass--metallicity relation can be reduced from 0.091 to 
0.077 dex by a three-dimensional relation between stellar mass, 
metallicity and 4000\AA\, break. The reduction of scatter in 
mass--metallicity relation suggests that the galaxy stellar age plays an 
important role as the second parameter in the mass--metallicity relation 
of LBAs. 
\end{abstract}


\section{Introduction}
Metallicity is a key parameter to probe the galaxy formation and evolution. Metal enrichment process of a galaxy involves many physical processes 
such as star formation, gas inflow and outflow. A correlation between metal abundance and galaxy total mass (gas and stellar mass) was found 
in the 1970s 
\citep{lequeux1979}; typically, the greater mass galaxies have,
the richer metallicity they are. Subsequently, \citet{tremonti2004} confirmed the relation at z $\sim$ 0.1 
by using $\sim$ 50000 galaxies from the Sloan Digital Sky Survey (SDSS; \citealt{abazajian2004}). 
They also showed that the mass--metallicity relation is tighter and more fundamental than the luminosity--metallicity relation. 
Recently, with 
deep survey of distant galaxies, many works investigate the mass--metallicity relation at intermediate 
\citep{maier2005,savaglio2005,zahid2011,perez-montero2013} and high redshifts 
\citep{erb2006, maiolino2008, mannucci2009, laskar2011,kulas2013,yuan2013}. 
The mass--metallicity relation at high redshift is shifted to lower metallicity compared to the local relation.

The scatter in the mass--metallicity relation is an important clue to understand the metal enrichment history of 
a galaxy. Many galaxy physical parameters are found to be correlated with the mass--metallicity relation, and likely to contribute to 
scatter in this relation. In the local Universe,
the most widely investigated parameter is star formation rate (SFR; \citealt{ellison2008,lara-lopez2010,mannucci2010,yates2012,
andrews2013}). \citet{mannucci2010} found that galaxies with higher SFR typically have lower metallicites at a fixed mass. 
They proposed that the scatter in mass--metallicity relation could be significantly reduced by the relation between stellar mass, 
metallicity and SFR (fundamental metallicity relation; FMR). They argued that the local FMR
does not evolve out to z $\sim$ 2.5. However, the FMR is found 
to be dependent on methodology \citep{yates2012,andrews2013} and aperture correction factor \citep{sanchez2013}. In addition to SFR, many
other works also suggest that the mass--metallicity relation is dependent on other physical parameters, such as optical colour 
\citep{tremonti2004}, galaxy size \citep{ellison2008}, morphology \citep{solalonso2010} or gas mass fraction \citep{hughes2013}. 
Studying the parameter dependence of mass--metallicity at high redshift is rather difficult because of limited size of the sample. 
Recently, \citet{yabe2013} and \citet{zahid2014a} severally obtained spectra of hundreds of 
star-forming galaxies (SFGs) at $z \sim$ 1.4--1.7 and analysed the parameter dependence of the mass--metallicity relation at this 
redshift range. However, with different sample sizes and parameter ranges, \citet{zahid2014a} found evidence for the dependence on SFR, 
while no clear trend with SFR was found in the sample in \citet{yabe2013}.

To study the star formation properties of high-redshift SFGs in more detail, \citet{heckman2005} and \citet{hoopes2007} 
selected a local population of UV bright compact galaxies, named as supercompact UV luminous galaxies (UVLGs) or Lyman-break 
analogues (LBAs). The LBAs resemble Lyman-break galaxies (LBGs) in many aspects, including luminosity, stellar mass, SFR, extinction, and metallicity \citep{basuzych2007,hoopes2007,buat2009,overzier2011,elmegreen2013}. 
\citet{hoopes2007} and \citet{overzier2010} compared the mass--metallicity relation of UV selected LBAs with the relation of local 
SFGs \citep{tremonti2004} and UV selected LBGs at $z\sim2.2$ \citep{erb2006}. They found that LBAs have similar metallicity to that 
of local SFGs at massive end ($>10^{10.5}M_{\odot}$) and high redshift LBGs at lower masses ($<10^{10.5}M_{\odot}$). The high-redshift 
LBGs are less metal enriched than the local SFGs by a factor of 2 around \citep{erb2006}.
Here, with a larger sample, 
we compare the mass--metallicity relation with high-redshift SFGs in more detail and analyze the parameter dependence of the relation of LBAs. 
As a good representative of high-redshift SFGs in local Universe, a study of the correlation between the gas metallicity and 
other galaxy properties of LBAs could improve our knowledge about the metal enrichment process in LBAs and high redshift SFGs.

Throughout this paper, we adopt the cosmological parameters with $H_0=70\, {\rm km s^{-1} Mpc}^{-1}$, $\Omega_{\Lambda}=0.73$ 
and $\Omega_{\rm m}=0.27$.

\section{Sample selection}
\subsection{Lyman-break analogues}
\citet{heckman2005} and \citet{hoopes2007} selected a local compact SFG population (i.e. supercompact UVLGs) which
is considered to be local analogues of LBGs. The selection criteria are based on far-ultraviolet (FUV) luminosity 
($L_{1530}>2\times10^{10}\ L_{\odot}$) and surface brightness ($I_{1530}>10^9\ L_{\odot}{\rm kpc}^{-2}$). Since FUV is sensitive 
to the dust attenuation, selection based on FUV may bias against dusty galaxies. Besides, $\ha$ is also an excellent index of star
formation and less sensitive to dust attenuation than FUV. Therefore, we obtain our sample by using $\ha$ luminosity and surface brightness from SDSS 
Data Release 10 (DR10; \citealt{Ahn2013}). We use the emission-line fluxes, stellar masses and 4000\AA\, break, $\dfn$, derived by the MPA-JHU group 
\citep{brinchmann2004}. A careful subtraction of stellar 
absorption-line spectrum is performed
before measuring the emission lines \citep{kauffmann2003}. As noted by \citet{groves2012}, the equivalent width (EW)
of $\hb$ in the MPA-JHU catalogue is underestimated by $\sim$ 0.35 ${\rm \AA}$. Since the minimum EW of $\hb$ of LBAs is 10.7 ${\rm \AA}$,  
this correction
is negligible.

The selection criteria for LBAs are (1) redshift $0.05<z<0.30$; (2) $\ha$ luminosity $L(\ha)>10^{41.8}\,{\rm erg\,s^{-1}}$; (3) 
$\ha$ surface brightness $I(\ha)>10^{40.5}\,{\rm erg\,s^{-1}\,kpc^{-2}}$; (4) $\dfn$ is less than 1.1; (5) the log stellar mass error 
 is less than 0.3 dex; 6) signal-to-noise (S/N) of (\ha) higher than 10. The lower redshift 
limit is set in order to ensure that the \oii$\lambda$3727 line is located in the spectral range of SDSS (3800--9200\AA). 
$\ha$ surface brightness $I(\ha)=L(\ha)/2{\rm \pi} r_{50,i}^2$, where $r_{50,i}$ is the half-light radius of objects in the SDSS 
$i$-band image. We use scale-length from the exponential model fit of the SDSS $i$-band image as the half-light radius at the $\ha$ band. 
We do not correct $\ha$ flux in our selection criteria for the intrinsic dust attenuation to be consistent with the uncorrected FUV criteria above.
The thresholds of $\ha$ luminosity and surface brightness are set to be comparable (i.e. with the same unobscured SFR)
with FUV criteria of supercompact UVLGs. 
We set a $\dfn$ upper limit to exclude intrinsic red (i.e. old) emission line galaxies.
Since no stellar mass error is available in the MPA-JHU catalogue, we defined the error as half of the difference between the 84th percentile 
and 16th percentile stellar mass probability distribution function (i.e. lgm\_{tot}\_{p84} and lgm\_tot\_p16 index) in the MPA-JHU catalogue. 
About 7 percent of galaxies have stellar mass error systematically
higher than the major sample and the dividing point is at $\sim 0.3$ dex. These galaxies also deviate the major sample in the mass--metallicity
diagram, which may be due to the poorly determined stellar mass.
The $\ha$ S/N criterion removes another six galaxies which do not have reliable detection of the $\ha$ line. 
In total, 817 galaxies in the SDSS DR10 are selected by our $\ha$ criteria.

Since fibre spectra only measure the centre regions of extended objects, aperture corrections are needed to get integrated 
properties. One simple estimate of aperture correction factor is the difference between fibre and total magnitude \citep{hopkins2003}. 
Owing to 
lack of knowledge about the spatial distribution of emission line, the correction factors for emission line are always estimated by the use of 
adjacent broad-band photometry. We use SDSS $i$-band photometry to roughly estimate the aperture correction factors. The median correction
factor of LBAs is only 1.5 
(0.18 dex), which indicates that two thirds of light is included in the SDSS 3-arcsec fibre. The aperture correction factor is 
rather small compared to normal SFGs because LBAs
are compact objects with typical sizes of $1\sim 2$ kpc \citep{overzier2009,overzier2010}. A comparison between the fibre SFR and
the aperture corrected SFR measurement in the MPA-JHU galSpecExtra catalogue 
\footnote{http://skyserver.sdss3.org/dr9/en/help/browser/browser.asp} (i.e. sft\_{tot}\_{p50} index) indicates smaller correction
factors with a median of 1.2 ( 0.07 dex). The aperture corrected SFR in MPA-JHU catalogue is derived from combining emission line 
within the fibre and aperture 
corrections by fitting model \citep{gallazzi2005,salim2007} into the photometry outside the fibre. With such high fibre covering fraction, 
the metallicity and dust extinction calculated with fibre spectra could represent the 
global values \citep{kewley2005}. The aperture correction method always depends on the assumption that the distributions of 
emission line (ionized gas emission) and broad-band photometry (stellar emission) are identical outside the fibre, which may introduce systematic
errors
in aperture correction. These systematic errors could introduce many extended normal SFGs 
which have more extended continuum emission than emission lines. Besides, the aperture correction factor for LBAs
is rather small and the majority of them could be selected by criteria within fibre. Therefore, 
we do not correct the aperture effect for $\ha$ flux in sample 
selection. However, we will use aperture-corrected SFR and gas mass fraction in subsequent 
analysis. 
% and aperture effect.

%AGN exclusion
The AGN contamination is excluded from our sample by the BPT diagram \citep{baldwin1981}. Fig. 1 shows the distribution of LBAs in 
the $\rm{log}(\nii/\ha)$ versus $\rm{log}(\oiii/\hb)$ diagram. We cross match the 817 galaxies with {\em GALEX} all sky survey catalogue
\citep{martin2005} and divide it into two subsamples. Galaxies that satisfy the FUV criteria of supercompact UVLGs are grouped into the 
`UV' selected subsample and others into the `non-UV' selected subsample. The red asterisks and blue circles represent the `non-UV'
and `UV' selected subsamples, 
respectively. 
It can be seen from Fig. 1 that the `non-UV' and `UV' LBAs show similar distribution of line ratios which suggests that they have similar
ionization properties. A two-sample Kolmogorov--Smirnov test suggests that, in $\rm{log}(\nii/\ha)$ and $\rm{log}(\oiii/\hb)$ 
distributions, the differences between two subsamples
are significant at a level less than 2$\sigma$. 
%between these two 
%subsamples in BPT diagrams. The null hypotheses that the two subsamples are drawn from different distributions of $\rm{log}(\nii/\ha)$ 
%and $\rm{log}(\oiii/\hb)$ is rejected at significance level of 0.059 and 0.168, 
The solid and dotted lines are the demarcation curves between SFGs and AGNs derived 
by \citet{kauffmann2003} and \citet{kewley2001}, respectively. Galaxies located between these two lines are often classified as 
composite objects which may host a mixture of AGN and star formation. We adopt the criteria of \citet{kauffmann2003} to remove AGNs.
The AGN fractions in the `UV' and `non-UV' subsample are 14 and 12 per cent, respectively. Finally, 703 galaxies (524 
`non-UV' and 179 `UV') 
remain and constitute our LBA sample, with a median redshift of $0.20\pm 0.04$.  

\begin{figure}
\centering
\includegraphics[width=9cm]{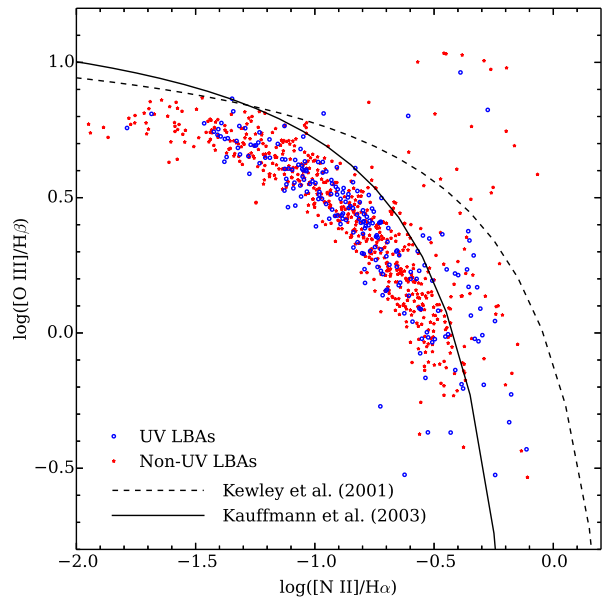}
%\resizebox{\hsize}{!}{\includegraphics{bpt-k.ps}}
\caption{BPT diagram for LBAs to exclude AGN. The red asterisks and blue circles are `non-UV' and `UV' subsamples, respectively. 
The solid and dotted lines are the separation lines for SFGs and AGNs, which are derived by \citet{kauffmann2003} 
and \citet{kewley2001}, respectively.}
\label{figure1}
\end{figure}

\subsection{SFGs in SDSS}

\begin{figure}
\centering
\includegraphics[width=9cm]{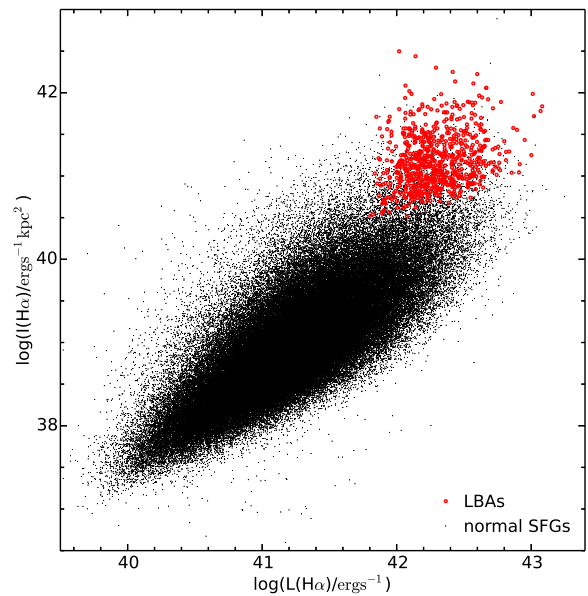}
%\resizebox{\hsize}{!}{\includegraphics{bpt-k.ps}}
\caption{$\ha$ surface bright versus $\ha$ luminosity. $\ha$ luminosity is corrected for dust extinction.
Stars are normal SFGs and red circles are LBAs.}
\label{figure2}
\end{figure}

To compare the parameter dependence of the mass--metallicity relation between
LBAs and local normal SFGs, we select an SFG sample from SDSS DR10 according to the selection criteria in 
\citet{mannucci2010}. The major selection criteria are (1) redshift between 0.07 and 0.3; (2) S/N ($\ha$)$ > 25$; 
(3) dust attenuation $\rm {A_v} < 2.5$.
Finally, 152477 galaxies remain in the normal SFG sample. 
Fig. 2 shows the distributions of LBAs and SFGs in the dust extinction-corrected $\ha$ luminosity versus $\ha$ surface brightness
diagram. It can be seen that LBAs are extreme cases of normal SFGs with high SFR and SFR surface densities.

\section{Mass--Metallicity relation}

There are several methods to determine the metallicity of an emission-line galaxy. The `$T_{\rm e}$-method', which is regarded as a
direct way to obtain metallicity, is based on the fact that metallicity is anticorrelated with the electron temperature $T_{\rm e}$. 
The electron temperature can be obtained using the ratio of the auroral to nebular emission lines of the same ion. Since the auroral 
lines are always very weak, many empirical and theoretical methods are developed (classified according to \citealt{kewley2008}). The 
empirical calibrations such as `N2 method' (\nii/$\ha$; \citealt{pp04}), `O3N2 method' ((\oiii/$\hb$)/(\nii/$\ha$); \citealt{pp04}) 
and `R23 method' ((\oii$\lambda$3727+\oiii$\lambda\lambda$4959,5007)/$\hb$; \citealt{p05}) are calibrated by the relation between
strong line ratios and the $T_{
\rm e}$-derived metallicities. The theoretical methods are also based on strong line ratios but calibrated by theoretical 
photoionization models. Although these methods are all based on strong line ratios, the metallicity derived from different methods can 
vary up to 0.7 dex \citep{shi2005, kewley2008}. The origin of the discrepancy between different
metallicity calibrations is not clear yet.

Obtaining spectra with wide wavelength range is difficult and time-consuming for high-redshift galaxies. Therefore the 
empirical N2 method, based on two adjacent lines (i.e. $\nii\lambda 6584$ and $\ha$), is widely used in high redshift works to obtain
metallicity. Another advantage of the N2 method is that it is insensitive to dust attenuation, because of the adjacency of 
$\nii\lambda 6584$ and $\ha$. The drawback of the N2 method is the saturation of N2 index above approximate solar abundance and dependence on
ionization parameters. 
With the significant discrepancies between different metallicity calibrations, comparison with other works should be very careful.
It is more straightforward to use the same metallicity calibration. Since one main goal of this paper is to compare the 
mass--metallicity relations of LBAs and high redshift SFGs, the empirical N2 method \citep{pp04} is used to determine metallicities
for our LBA sample. 

\citet{pp04} obtained metallicities of 137 $\hii$ regions using $T_{\rm e}$ method or detailed photoionization modelling when the direct
method cannot be applied. The linear calibration of the N2 method from \citet{pp04} is given as
\begin{equation}
12+{\rm log(O/H)}=8.90+0.57\times {\rm N2} ,
\end{equation}
where N2 is defined as ${\rm N2}={\rm log}(\nii\lambda 6584/\ha)$. The calibration is valid for $-2.5 < N2 < -0.3$ and the intrinsic
dispersion is 0.18 dex. 

The left-hand panel of Fig. 3 shows the distribution of LBAs in the mass--metallicity diagram. Galaxies in the UV subsample are represented
as blue circles and those in the non-UV subsample as red asterisks.
It can be seen that these two subsamples have a similar distribution
in terms of mass and metallicity. A two-sample Kolmogorov--Smirnov test suggests that, in mass and metallicity  
distributions, differences
are significant at a less than 2$\sigma$ level. The agreement between the UV and non-UV subsample suggests that 
the deviation of mass--metallicity relation between UV selected LBAs and local SFGs, 
which was reported 
by \citet{hoopes2007},
could not be explained by selection bias against dusty objects. The deviation may be due to the high SFR and $\dfn$ of LBAs compared to 
normal SFGs, which will be discussed in Section 4.

%A two-sample Kolmogorov-Smirnov test suggest that there is no significant difference between these
%two subsamples in the mass metallicity relation. The null hypothesis is rejected at level of 0.059 and 0.282 for metallicity and 
%mass distributions, respectively. 

The data points in Fig. 3 are divided into six bins with binsize of 0.4 dex in mass. The square points are the median value in each bin. The
vertical and horizontal errorbars are the median absolute deviation from median of metallicity and mass in each bin, respectively. 
There is a positive correlation between the two quantities and the data points are fitted in a logarithmic form defined as 
\begin{equation}
12+{\rm log(O/H)}=Z_0-{\rm log}[1+(M_*/M_0)^{-\gamma}]
\end{equation}
\citep{moustakas2011}. The best-fitting parameters are $Z_0=8.68$, $M_0=9.63$ and $\gamma=0.74$. The metallicity residuals 
from the 
logarithmic fit are showed in the insert panel of Fig. 3. The Spearman rank test gives a correlation coefficient of -0.004, which
indicates that the fitted mass--metallicity relation does not introduce artificial trends across the mass
range.

\begin{figure*}%[!tp]
\centering
\includegraphics[width=15cm]{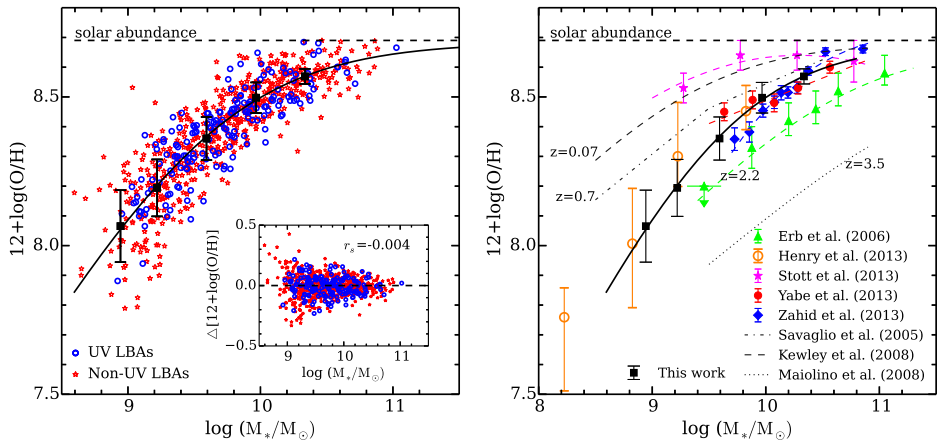}         
%\resizebox{\hsize}{!}{\includegraphics{mass-metal-n2-k.ps}}
\caption{Left-hand panel: the mass--metallicity diagram for galaxies in UV subsample (blue circles) and non-UV subsample (red asterisks). 
The solid black curve is the best-fitting line and the black squares are binned data. Error bars are the median absolute
deviation from median in each bin. The bottom-right inset panel is the metallicity residuals from the logarithmic fit. Right-hand panel: 
comparison of mass--metallicity relation of LBAs with the relations at high redshifts. The horizontal dashed line is solar metallicity
(12+log(O/H)=8.69; \citealt{asplund2009}).}
\label{figure3}
\end{figure*}

The 
mass--metallicity relation at high redshift has been largely exploited in recent years ($z$=0.84 and 1.47, \citealt{stott2013}; 
$z\sim1.4$, \citealt{yabe2013}; $z\sim1.6$, \citealt{zahid2014a,zahid2014b};
$z\sim1.76$, \citealt{henry2013}; $z\sim2$, \citealt{erb2006} and \citealt{hayashi2009}; $z\sim3.5$, \citealt{maiolino2008}).
The comparison
of the mass--metallicity relation between LBAs and these high-redshift SFG samples 
is presented in the right-hand panel of Fig. 3. 
The local SFGs sample ($z\sim0.07$; \citealt{kewley2008}) is also included as reference. For a fair comparison, the masses are corrected to
Kroupa initial mass function \citep{kroupa2001} based stellar masses and metallicities to the empirical N2 method \citep{pp04} through the conversion in 
\citet{maiolino2008}.

It is worth noticing that the mass--metallicity relation of LBAs is in remarkable consistency with that at $z\sim$ 1.4--1.7 
\citep{yabe2013,zahid2014a}. The consistency is extended to lower masses by \citet{henry2013}, who investigated the mass--metallicity relation 
of low mass galaxies at $z\sim1.76$ at mass down to $\sim10^{8.5}M_{\odot}$. \citet{stott2013}
measured the mass--metallicity relation for a combined sample of galaxies at $z\ =\ 0.84$ and 1.47. However, they found that the 
mass--metallicity relation at $z\ =\ 0.84$ and 1.47 is comparable to the local mass--metallicity relation and concluded that there is no 
evolution of the mass--metallicity relation with redshift. They argued that the origin of evolution in the mass--metallicity relation 
is selection bias in galaxies with bright UV luminosity and high SFR which tend to be more metal poor than normal galaxies.
However, \citet{zahid2014a} selected an SFG sample at $z\sim1.6$ based on $sBzK$ selection \citep{daddi2004}, which does not suffer the selection
bias mentioned in \citet{stott2013}, and argued that the evolution of mass--metallicity relation seems to be real.

It can be seen in Fig. 3 that massive LBAs have metallicities similar to that of normal SFGs at $z\sim0.07$. However, at lower 
masses, the metallicity discrepancy between LBAs and local SFGs becomes larger, which is consistent with the deviation 
found in \citet{hoopes2007}. 
\citet{savaglio2005} measured the mass--metallicity 
relation of SFGs at $z\sim0.7$ (dot--dashed line in the right-hand panel of Fig. 3) and interpreted the flattening of mass--metallicity 
relation from $z\sim0.7$ to $z\sim0.1$ as a result of downsizing evolution of galaxy population. Massive galaxies at high redshift
reach high metallicity, while low-mass systems enrich their interstellar medium on long time-scales and are still converting gas into stars at present
time. In terms of downsizing scenario, the local LBAs are less evolved compared to the local SFGs. 
It is also interesting to note that the mass--metallicity relation at $z \sim$ 1.4--1.7 \citep{yabe2013,zahid2014a}, supplemented by the
low mass galaxies from \citet{henry2013}, is roughly parallel to that at $z\sim2.2$ \citep{erb2006} and $z\sim3.5$ 
\citep{maiolino2008}. The shape of mass--metallicity relation does not significantly change from $z\sim3.5$ to $z \sim$ 1.4--1.7, which
suggests a mass-independent metal enrichment process. This is quite different from the downsizing/mass-dependent evolution at lower
redshift. However, a detailed consideration of sample selection biases at each epoch and larger sample at high redshift with 
sufficient low-mass galaxies is required to confirm or reject this trend.

\section{Dependence of the mass--metallicity relation on galaxy properties}

\begin{figure}%[!tp]
\centering
\includegraphics[width=9cm]{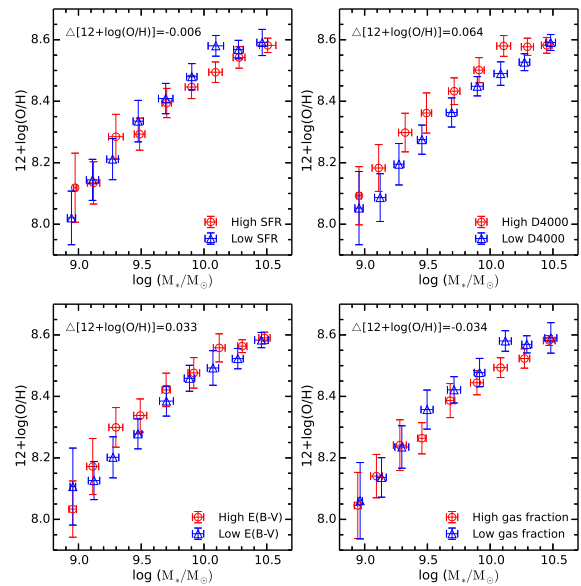}
%\resizebox{\hsize}{!}{\includegraphics{mass-metal-bin-n2-median-k.ps}}
\caption{The dependence of the mass--metallicity relation on the galaxy physical parameters, including SFR, $\dfn$, $E(B-V)$ and
gas mass fraction. In each mass bin, we divide galaxies into two groups by the parameter. We calculate the median of mass and 
metallicity for each group in each bin. The median absolute deviation from median value is used to estimate the spread and represented
as errorbar. The deviation shown in the head of each panel represents the average offset in each bin.} 
\label{figure4}

\end{figure}

The dependence of the mass--metallicity relation on galaxy properties has been explored by many works (\citealt{ellison2008}, 
\citealt{mannucci2010}, \citealt{lara-lopez2010}, \citealt{yates2012}, \citealt{sanchez2013} and \citealt{andrews2013} for local SFGs;
\citealt{stott2013}, \citealt{yabe2013} and \citealt{zahid2014a} for high-redshift SFGs). Here, we examine that
for LBAs and compare with the parameter dependence of the mass--metallicity relation at
high redshift. The parameters examined are the intrinsic SFR, $\dfn$, $E(B-V)$ and gas mass fraction. The intrinsic SFR is derived 
from $\ha$ luminosity by use of the relation from \citet{kennicutt1998}. We correct $\ha$ luminosity for dust extinction by using
the Balmer decrement method and adopting the Calzetti extinction law \citep{calzetti2000}. We also account for the aperture effect using
the correction method mentioned in Section 2.1. The obtained SFR ranges from 4 to 100
$M_{\odot}{\rm yr}^{-1}$ with a median of $15.9\pm5.7$ $M_{\odot}{\rm yr}^{-1}$. The corrected SFR is consistent with the aperture-corrected
SFR measurement in the MPA-JHU catalogue. The $E(B-V)$, which is derived from the Balmer 
decrement method with
Calzetti extinction law, ranges from 0.0 to 0.49 mag with a median of $0.17\pm0.06$. The $\dfn$ is taken from the `d4000\_{n}' index in the 
MPA-JHU galSpecIndx catalogue \footnote{http://skyserver.sdss3.org/dr9/en/help/browser/browser.asp} \citep{brinchmann2004} and calculated
according to the \citet{balogh1998} definition. The LBA sample has $\dfn$ value ranging from 0.45 to 1.1.
To accurately estimate the gas mass, CO measurements, which is extremely time-consuming 
for a large sample of normal SFGs, are necessary. Alternatively, 
a correlation between surface densities of gas mass and star formation was found and independent of galaxy type
(i.e. the Schmidt law; 
\citealt{kennicutt1998}). Assuming such a law is suitable for LBAs, we could roughly estimate the gas mass from dust and aperture corrected 
$\ha$ luminosity and size of galaxies. 
This indirect method is also used by \citet{tremonti2004} and \citet{erb2006} to derive gas mass fractions.
The size of galaxy, which represents the size of the emission region 
of stellar component, is assumed to be the same to that of $\ha$ emission region. The obtained gas mass fraction 
, $M_{\rm gas}/(M_{\rm gas}+M_{*})$, ranges from 0.01 to 0.79 with a median of $0.35\pm0.14$.

\begin{table*}
  \caption{Parameter dependence of the mass--metallicity relation. For each parameter and stellar mass bin, the median values
are listed.}
  \label{table:1}
%\centering
  \begin{tabular}{l|cccc|cccc|cccc|cccc}
\hline\hline
       ID      &  log($M_*$/${\rm M}_{\odot}$) & 12+log(O/H) [N2] ([N2+R23])  & SFR (${\rm M}_{\odot}{\rm yr}^{-1}$) & 
       $\dfn$ & $E(B-V)$ & Gas fraction & $N\ $ $^a$   \\
\hline
1  & 8.96$\pm 0.03$ & 8.06$\pm 0.12$ (8.15$\pm 0.14$) & 8.3$\pm 2.2$  & 0.69$\pm 0.04$ & 0.08$\pm 0.04$ & 0.61$\pm 0.05$ & 25 (24) \\
2  & 9.12$\pm 0.05$ & 8.14$\pm 0.07$ (8.23$\pm 0.10$) & 9.7$\pm 2.4$ & 0.77$\pm 0.07$ & 0.08$\pm 0.05$ & 0.55$\pm 0.06$ & 68 (61)\\
3  & 9.28$\pm 0.05$ & 8.24$\pm 0.07$ (8.39$\pm 0.10$) & 13.2$\pm 3.4$ & 0.85$\pm 0.06$ & 0.11$\pm 0.04$ & 0.52$\pm 0.07$ & 98 (90) \\
4  & 9.48$\pm 0.04$ & 8.31$\pm 0.06$ (8.47$\pm 0.08$) & 15.0$\pm 5.4$ & 0.91$\pm 0.06$ & 0.13$\pm 0.04$ & 0.46$\pm 0.08$ & 92 (86)\\
5  & 9.70$\pm 0.04$ & 8.40$\pm 0.05$ (8.60$\pm 0.07$) & 15.6$\pm 4.2$ & 0.98$\pm 0.04$ & 0.17$\pm 0.04$ & 0.36$\pm 0.06$ & 99 (97)\\
6  & 9.90$\pm 0.05$ & 8.47$\pm 0.04$ (8.69$\pm 0.06$) & 16.5$\pm 4.9$ & 1.02$\pm 0.03$ & 0.20$\pm 0.05$ & 0.29$\pm 0.07$ & 117 (115)\\
7  &10.09$\pm 0.06$ & 8.53$\pm 0.05$ (8.77$\pm 0.08$) & 19.3$\pm 5.1$ & 1.05$\pm 0.03$ & 0.23$\pm 0.05$ & 0.22$\pm 0.05$ & 82 (80)\\
8  &10.28$\pm 0.05$ & 8.55$\pm 0.04$ (8.80$\pm 0.05$) & 27.9$\pm 9.7$ & 1.05$\pm 0.03$ & 0.27$\pm 0.04$ & 0.20$\pm 0.05$ & 59 (59)\\
9  &10.48$\pm 0.05$ & 8.59$\pm 0.02$ (8.85$\pm 0.05$) & 34.0$\pm 7.8$& 1.07$\pm 0.02$ & 0.30$\pm 0.05$ & 0.16$\pm 0.03$ & 34 (33)\\
%10 &10.67$\pm 0.06$ & 8.59$\pm 0.04$ (8.95$\pm 0.04$) & 30.1$\pm 13.0$& 1.05$\pm 0.02$ & 0.27$\pm 0.08$ & 0.11$\pm 0.03$ & 23 (23)\\
\hline

\end{tabular}\\
Note:
$^a\ $ Number of galaxies in each stellar mass bin. The value in parentheses is presented for LBAs which have metallicity determined 
by using N2+R23 calibration.
\end{table*}

\begin{figure}%[!tp]
\centering
\includegraphics[width=9cm]{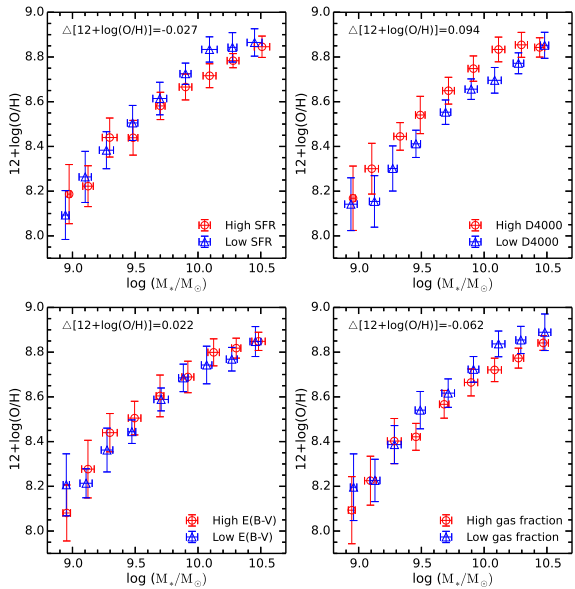}
%\resizebox{\hsize}{!}{\includegraphics{mass-metal-bin-r23-median-k.ps}}
\caption{The same diagram with Fig. 4, except the metallicity is calculated using the method combining R23 and N2 calibration from 
\citet{maiolino2008}.}
\label{figure5}
\end{figure}

%\begin{table*}
%\caption{The same as Table 1 but metallicity is determined using $R_{23}$ method.}
%\label{table:2}
%\begin{tabular}{l c c c c c c cr}
%\hline\hline
%       ID      &  log($M_*$/${\rm M}_{\odot}$) & 12+log(O/H) [$R_{23}$]  & SFR (${\rm M}_{\odot}{\rm yr}^{-1}$) & 
%       $\dfn$ & $\sl E(B-V)$ & gas fraction & $N$   \\
%\hline
%1  & 9.12$\pm 0.05$ & 8.43$\pm 0.09$ &  9.4$\pm 2.0$ & 0.79$\pm 0.04$ & 0.08$\pm 0.04$ & 0.55$\pm 0.05$ & 53 \\
%2  & 9.30$\pm 0.06$ & 8.55$\pm 0.09$ & 13.2$\pm 3.5$ & 0.86$\pm 0.05$ & 0.11$\pm 0.04$ & 0.52$\pm 0.06$ & 87 \\
%3  & 9.48$\pm 0.04$ & 8.62$\pm 0.08$ & 14.4$\pm 5.0$ & 0.91$\pm 0.05$ & 0.13$\pm 0.04$ & 0.45$\pm 0.08$ & 93 \\
%4  & 9.70$\pm 0.04$ & 8.73$\pm 0.06$ & 15.6$\pm 4.2$ & 0.98$\pm 0.04$ & 0.17$\pm 0.04$ & 0.36$\pm 0.06$ & 99 \\
%5  & 9.90$\pm 0.05$ & 8.81$\pm 0.07$ & 16.4$\pm 4.5$ & 1.02$\pm 0.03$ & 0.20$\pm 0.04$ & 0.28$\pm 0.07$ & 122 \\
%6  &10.09$\pm 0.06$ & 8.88$\pm 0.08$ & 19.0$\pm 5.7$ & 1.05$\pm 0.04$ & 0.22$\pm 0.05$ & 0.21$\pm 0.05$ & 88 \\
%7  &10.29$\pm 0.05$ & 8.90$\pm 0.08$ & 25.3$\pm 8.7$ & 1.04$\pm 0.03$ & 0.26$\pm 0.05$ & 0.18$\pm 0.05$ & 71 \\
%8  &10.49$\pm 0.05$ & 8.93$\pm 0.07$ & 26.7$\pm 12.1$& 1.06$\pm 0.02$ & 0.25$\pm 0.06$ & 0.13$\pm 0.04$ & 49 \\
%9  &10.67$\pm 0.06$ & 8.95$\pm 0.04$ & 30.1$\pm 13.0$& 1.05$\pm 0.02$ & 0.27$\pm 0.08$ & 0.11$\pm 0.03$ & 23 \\
%\hline
%\end{tabular}\\
%\end{table*}

To explore the effect of different metallicity calibration methods on the parameter dependence of the mass--metallicity relation, 
we also use the method given in \citet{mannucci2010} to determine the metallicities of LBAs and normal SFGs. Two independent calibrations
based on the N2 and R23 ratio
as described in \citet{maiolino2008} are used (hereafter N2+R23 calibration). When both two calibrations are available
(i.e. log($\nii/\ha$) $ <$ --0.35 and log(R23) $ <$ 0.9) 
and the difference between the two derived metallicites is less than 0.25 dex, the average of the two values is the final metallicity. It is
suggested that the R23 calibration has double-valued solutions. Since the $\nii/\oii$ ratio is a strong function of metallicity \citep{kewley2002},
we use the $\nii/\oii$ ratio to break the R23 degeneracy. We take the higher metallicity value for galaxies with log($\nii/\oii$) $> -1.2$ 
and lower value for 
galaxies with log($\nii/\oii$) $ < -1.2$.

The LBA 
sample is divided into 12 mass bins with binsize of 0.2 dex. To study the parameter dependence of the mass--metallicity relation, we divide
the objects in each mass bin into two groups according to the median value of the parameter in the mass bin. 
Metallicity in Figs 4 and 5 is calculated using empirical N2 method and N2+R23 calibrations, respectively.
The circles and triangles in Figs 4 and 5 
represent the median of mass and metallicity for the group with high and low value, respectively. The median absolute
deviation from median value is used to estimate the spread of $x$- and $y$-axes and represented as errorbar in Figs 4 and 5. 
Three bins in Figs 4 and 5, which have objects less than 20, are excluded for analysis. The median value of each parameter
in each stellar mass bin is listed in Table 1.
%for metallicity determined using N2 method and Table 2 for metallicity determined
%using $R_{23}$ method.

\subsection{SFR}
There is no systematical offset between high-SFR and low-SFR group galaxies in Fig. 4. However, systematic deviation seems to exist 
in intermediate-mass bins in Fig. 5. The relation between mass, metallicity and SFR in local SFGs has been discovered and studied in many 
recent works 
\citep{ellison2008,lara-lopez2010,mannucci2010,yates2012,andrews2013,sanchez2013,stott2013}. Galaxies with higher SFR tend to have lower 
metallicities at a fixed stellar mass \citep{ellison2008,mannucci2010,andrews2013}. Although FMR is found in local Universe, 
the relation between stellar mass, metallicity and SFR in high-redshift galaxies is not 
very clear yet. The mass--metallicity relation is dependent on SFR in the sample of \citet{zahid2014a} at $z\sim 1.6$. However,
\citet{yabe2013} did not find systematical trend that galaxies with higher SFR have lower metallicity in their SFG sample
at $z\sim 1.4$.

Based on the local relation between stellar mass, 
metallicity and SFR, \citet{mannucci2010} proposed that the scatter in the mass--metallicity relation could be significantly reduced by 
including SFR as another free parameter. The three-dimensional relation is named as the `fundamental metallicity relation' (FMR). 
They defined a new quantity $\mu_{\alpha}$, which is a combination of stellar mass and SFR 
($\mu_{\alpha}={\rm log}(M_*/M_{\odot})-\alpha {\rm log(SFR)}$). They found that $\alpha=0.32$ minimizes the scatter in the local
FMR and the value does not change in the FMR of SFGs at $z<2.5$. To verify whether our
LBA sample is consistent with the FMR, we plot the FMR for LBAs in Fig. 6. 
The solid curve is the FMR in \citet{mannucci2010}, which is not corrected for the aperture effect.
We also calculated the FMR of normal SFGs, shown as dashed curve, 
using aperture 
corrected SFR
from the MPA-JHU catalogue. It can be seen that the FMR of normal SFGs with aperture-corrected SFR shifts left
to the original FMR in \citet{mannucci2010} with fibre SFR, which is consistent with the estimate of FMR shift in \citet{zahid2014a}.
The FMR of LBAs is roughly consistent with both two FMRs of local normal SFGs. We note that there is better agreement between local FMR and 
the relation of LBAs if we use SFRs without aperture corrections. However, the FMR connects global properties of a galaxy and 
SDSS fibre SFR is not an appropriate measurement to derive such a relation.
Since the mass--metallicity relation of LBAs is not strongly dependent on SFR, an
FMR that combines stellar mass and SFR does not reduce the scatter in the mass--metallicity relation
notably. It should be noted that LBAs are subsample of normal SFGs with high SFR. The slight SFR dependence
may be due to the relative narrow SFR range of LBA sample.

\begin{figure}%[!tp]
\centering
\includegraphics[width=9cm]{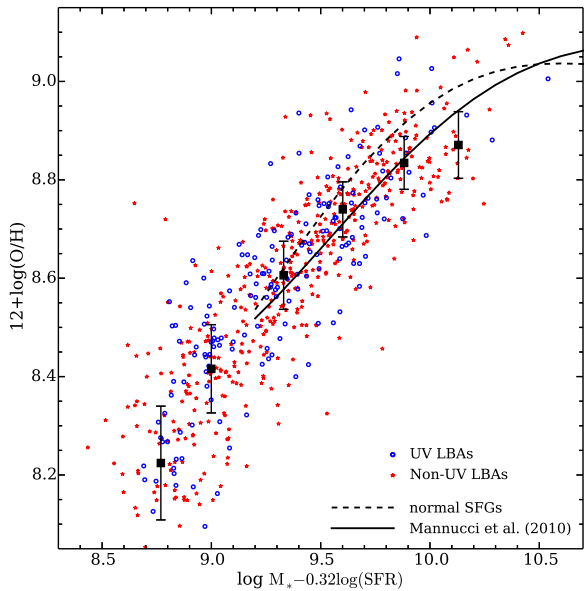}
%\resizebox{\hsize}{!}{\includegraphics{mass-metal-sfr-fmr.eps}}
\caption{The FMR of our LBA sample. The metallicity is calculated by the same calibration in 
\citet{mannucci2010}. The binned data points and errorbars are calculated by the same procedure as mentioned in Fig. 4. The solid
line is the metallicity--$\mu_{0.32}$ relation for the local SFG sample by \citet{mannucci2010}. The dashed curve is the FMR of normal SFGs using 
aperture-corrected SFR.}
\label{figure6}
\end{figure}

\begin{figure*}%[!tp]
\centering
%\resizebox{\hsize}{!}{\includegraphics{fmr-d4000.eps}}
\includegraphics[width=20cm]{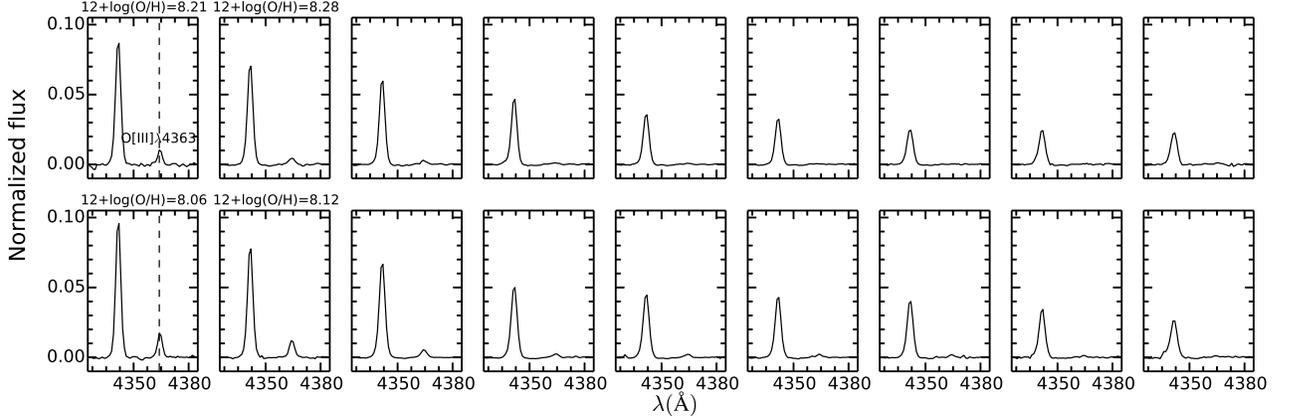}
\caption{Comparison of stacked spectra between high and low $\dfn$ groups. From left to right, the nine columns are stacked spectra of
the nine stellar mass bins 
from $10^9$ to $10^{11} {\rm M_{odot}}$. High- and low-$\dfn$ groups are displayed at upper and lower rows. The dashed line in each
row marks the \oiii$\lambda$ 4363 line. The title of the first two columns shows the metallicities calculated by using the $T_{\rm e}$-method.}
\label{figure7}
\end{figure*}

\begin{figure*}%[!tp]
\centering
%\resizebox{\hsize}{!}{\includegraphics{fmr-d4000.eps}}
\includegraphics[width=15cm]{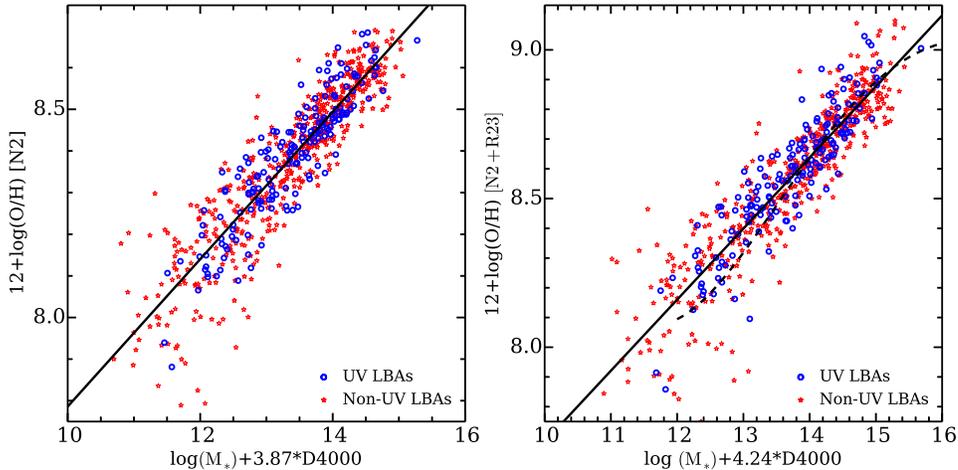}
\caption{Metallicity as a function of $\phi_{\alpha}$. Left-hand panel is for the empirical N2 method and right-hand panel for the method combining N2
and R23
calibrations. Solid lines are best-fitting metallicity--$\phi_{\alpha}$ relations. The dashed curve in the right panel is the 
metallicity--$\phi_{\alpha}$ relation of normal SFGs.}
\label{figure8}
\end{figure*}

\subsection{$\dfn$}
\citet{yabe2013} found that the metallicity of galaxies at high redshift correlates with the rest-frame NUV$-$optical colour at
a fixed mass. Galaxies with redder colour tend to have higher metallicities. \citet{hayashi2009} suggested that the difference in
observed $R-K$ colour, which covers the 4000 ${\rm \AA}$ break, may contribute to the discrepancy between the mass--metallicity
relation they derived 
at z $\sim$ 2 and that from \citet{erb2006}. It can be seen from the upper-right panel of Fig. 4 that there is strong 
correlation between the mass--metallicity relation and $\dfn$ for our LBA sample. The trend could also be seen from Fig. 5 which suggests that 
the correlation with $\dfn$ is not dependent on the metallicity methodology. It is worth noting that the average offset between high- and 
low-$\dfn$ groups is as large as 0.06 dex, which is the largest deviation among the four galaxy physical parameters. 
Since the metallicity calibrations based on strong line ratio have large uncertainties,
we also use the $T_{\rm e}$-method to compare the metallicity of high- and low-$\dfn$ groups. 
We split our sample into nine stellar mass bins from $10^9$ to $10^{11} {\rm M_{\odot}}$. Each size is 0.2 dex. 
Each stellar 
mass bin is then divided into two groups according to the median value of $\dfn$.
Then we 
follow the spectra stack and continuum subtraction steps of \citet{andrews2013}. Fig. 7 shows the reduced restframe 
spectra of 4250--4450 ${\rm \AA}$
of each group in each mass bin. The spectra have been normalized to dust-corrected $\hb$ flux.
The high-$\dfn$ groups are shown in upper row. In each mass bin, it can be seen that the \oiii$\lambda$4363
auroral line is much stronger in the low-$\dfn$
groups. It is suggested that the strength of \oiii$\lambda$4363 line is strongly correlated with the electron temperature and anti-correlated with metallicity. Therefore metallicity based on the $T_{\rm e}$-method should be higher in high-$\dfn$ groups. Since the \oiii$\lambda$4363 line 
is too weak in most 
mass bins to be reliably measured, we calculate the metallicity of the first two mass bins using the $T_{\rm e}$-method 
reported by \citet{izotov2006}.
The results are shown as title in the corresponding panel. The metallicity difference in the first two mass bins 
between high-$\dfn$ and low-$\dfn$ groups is $\sim$ 0.15 dex, which is comparable to the value in Figs 4 and 5.

Following the 
definition of $\mu_{\alpha}$ in \citet{mannucci2010}, we introduce a new quantity $\phi_{\alpha}$ as a combination of stellar mass
and $\dfn$: $\phi_{\alpha}={\rm log}M_{\odot}+\alpha*D_n(4000)$. The original residual in the mass--metallicity relation with N2 method 
is 0.091 dex and can be reduced to 0.077 dex in the metallicity--$\phi_{\alpha}$ relation with best-fitting $\alpha=3.87$. 
The reduction
of residual in N2+R23 calibrated metallicity is 0.03 dex, from 0.13 dex to 0.10 dex with best fitted $\alpha=4.24$. This is the
first time that the $\dfn$ is found to be one of the contribution to the scatter in the mass--metallicity relation. The 
projection of the three-dimensional relation between stellar mass, metallicity and $\dfn$ on metallicity-$\phi_{\alpha}$ surface is
shown in Fig. 8. The left-hand panel is for empirical N2 method and the right-hand panel for N2+R23 method calculated metallicity. 
The solid lines show the
best-fitting metallicity--$\phi_{\alpha}$ relation as follows:
\begin{eqnarray}
12+{\rm log(O/H)}\ \ [\rm{N2}] & = & 6.016 +0.177\times \phi_{3.87}  \\
12+{\rm log(O/H)}\ [N2+\rm{R23}]  & = & 5.292 +0.239 \times \phi_{4.24} .
\end{eqnarray}
We also find a three dimensional relation between mass, metallicity and $\dfn$ of local normal SFGs with moderate reduction 
of residual metallicity 
scatter ($\sim$ 0.01 dex). The projection of this relation is shown as dashed curve in the right-hand panel of Fig. 8. There is good agreement 
between the 
relation of LBAs and that of normal SFGs, which suggests a universal relation of different types of SFGs. Furthermore, the deviation 
of mass--metallicity relation between LBAs and normal SFGs could also be partially explained by their different stellar age. Normal SFGs are 
typically more evolved and thus have higher metallicities than LBAs.

Since $\dfn$ is considered as a good indicator of galaxy stellar age \citep{kauffmann2003}, the dependence on $\dfn$ could be 
interpreted as: galaxies with elder stellar age always have higher metallicity at a fixed stellar mass.  This trend seems to be
understandable. Galaxies with the same current stellar mass did not produce the same amount of stellar mass
along with their formation history. Actually, more stars in elder galaxies have left the stellar main sequence than those in younger 
galaxies with the same current stellar mass, in average. Therefore, the integral of stellar mass along with the lifetime of a galaxy should 
be higher for elder galaxies. Thus elder galaxies experience more metal enrichment process 
(such as supernova and/or massive stellar wind) and have relative higher metallicities. Given this interpretation, 
the dependence on stellar age of the mass--metallicity 
relation is a natural result of passive evolution since one can only measure the current galaxy mass.
%{\bf The slighter reduction of 
%residual metallicity scatter of SFGs compared with LBAs suggests that the stellar age (i.e. passive evolution)
%may play a more important role in the metal enrichment
%history of LBAs.}
%The notable reduction of scatter in metallicity 
%confirms that the galaxy stellar age plays as an important second parameter in the mass--metallicity relation.
%evolution stage information compared to metallicity. 

\subsection{$E(B-V)$}
There is a slight trend in the bottom-left panel of Fig. 4 that galaxies with higher $E(B-V)$ tend to have higher 
metallicities. This trend seems to be weaker in Fig. 5 with metallicity determined using the N2+R23 method. The average offset in each 
mass bin between large and small $E(B-V)$ is $\sim$ 0.03 dex. Similar trends are observed in a high-redshift galaxy sample at $z \sim$
1.4--1.7 \citep{yabe2013,zahid2014a}. This trend could be comprehensible because $E(B-V)$ reflects the dust extinction, while the dust 
content in a galaxy is strongly correlated with metal abundance in the galaxy \citep{heckman1998,reddy2010}. The average dependence
on $E(B-V)$, which is $\Delta[12+{\rm log(O/H)}]/\Delta E(B-V)=0.37\ {\rm dex\, mag}^{-1}$, is comparable with the $E(B-V)$ dependence
of galaxies at $z \sim$ 1.4 ($0.56 \  {\rm dex\, mag}^{-1}$) from \citet{yabe2013}.

\subsection{$Gas fraction$}
It can be seen in bottom right of Fig. 4 that the deviation between two group galaxies is slight and they are almost 
on the same mass--metallicity 
relation which resembles the behaviour of galaxy sample at $z \sim$ 1.4 \citep{yabe2013}. It is noted by \citet{yabe2013} that the 
dependence on gas mass fraction seems due to the mass--metallicity relation itself since the gas mass fraction correlates 
strongly with the stellar mass. It is worth noting that the dependence on gas mass fraction of mass--metallicity relation may be 
reflected by the dependence on SFR, since the estimation of gas mass is closely connected to the SFR.
\citet{hughes2013} found that the mass--metallicity relation is dependent on $\hi$ gas mass fraction; galaxies
with higher gas fractions have lower oxygen abundances at a fixed mass. 
With limited data set of CO maps and large uncertainties in CO-to-${\rm H}_2$
conversion factor, the dependence on ${\rm H}_2$ gas mass fraction is not clear yet.

\section{Summary}
The main goal of this work is to study the mass--metallicity relation of LBAs and the dependence of this relation on galaxy
properties. Since $\ha$ suffers less dust attenuation than UV photons, we select LBA sample according to
$\ha$ luminosity and surface brightness criteria to obtain a larger LBA sample with less bias. In total, 703 galaxies 
are selected from the SDSS DR10 as our LBA sample, and only $\sim$ 25 per cent of them can be recognized by the previous FUV 
selection criteria. The main results and conclusion of this paper are as follows.

\begin{itemize}
 \item The mass--metallicity relation of LBAs is in good agreement with that of SFGs at $z \sim$ 1.4--1.7 in stellar mass 
range of $10^{8.5}M_{\odot}<M_{*}<10^{11}M_{\odot}$. 

 \item The mass--metallicity relation is found to be strongly correlated with $\dfn$; galaxies with higher 
$\dfn$ typically have higher metallicity at a fixed mass. This trend is independent of the methodology of metallicity. 
We confirm the dependence on $\dfn$ by using the $T_{\rm e}$-method for estimating metallicity of stacked spectra.
A correlation of mass--metallicity relation with 
NUV$-$optical colour is probably due to the correlation with $\dfn$.
About 17 per cent of the total scatter in the mass--metallicity relation is due to the systematic effect with $\dfn$.
This suggests that the 
galaxy stellar age plays an important role as a second parameter in the mass--metallicity relation of LBAs and even high redshift SFGs.

\item The mass--metallicity relation of local normal SFGs is also found to be correlated with $\dfn$. There is good agreement between
the three dimensional relation of mass, metallicity and $\dfn$ of normal SFGs and LBAs, 
which suggests a universal relation of different type of SFGs.

 \item There is no strong correlation between mass--metallicity relation of LBAs and SFR.
The relation between stellar mass, metallicity and SFR of LBAs is roughly consistent
with the local FMR. The lack of dependence on SFR may be due to the narrow range of SFR of LBA sample.
%The SFR, which is not corrected for aperture effect in the local FMR, probably causes the deviation. 
 
 \item At a fixed mass, dustier galaxies typically have higher metallicities. This trend is consistent with SFGs at high-redshift 
\citep{yabe2013,zahid2014a}.
\end{itemize}

The similarity of mass--metallicity relation and its dependence on galaxy properties between LBAs and high redshift SFGs makes
it more explicit that local LBAs are located at a similar evolution stage to SFGs at $z \sim$ 1.4--1.7. 

\section*{Acknowledgements}

We are grateful to referee and Roderik A. Overzier for all insightful suggestions 
and comments and the MPA-JHU teams for their public measured quantities 
on SDSS galaxies. This work is supported by the National Natural Science Foundation of China (NSFC, Nos. 11225315, 1320101002, 11433005, and 11421303), the Strategic Priority Research Program "The Emergence of Cosmological Structures" of the Chinese Academy of Sciences (No. XDB09000000), the Specialized Research Fund for the Doctoral Program of Higher Education (SRFDP, No. 20123402110037), and the Chinese National 973 Fundamental Science Programs (973 program) (2015CB857004).

\end{document}